\documentclass[fleqn,usenatbib,letters]{mnras}

\usepackage{newtxtext}
\usepackage[varvw]{newtxmath}

\usepackage[T1]{fontenc}

\DeclareRobustCommand{\VAN}[3]{#2}
\let\VANthebibliography\thebibliography
\def\thebibliography{\DeclareRobustCommand{\VAN}[3]{##3}\VANthebibliography}

\usepackage{graphicx}	
\usepackage{amsmath}	
\usepackage{multirow}
\usepackage{verbatim}

\title[Demystifying shock breakout spectra]{Demystifying shock breakout spectra}

\author[Irwin \& Hotokezaka]{
Christopher M. Irwin,$^{1}$\thanks{E-mail: irwincm@g.ecc.u-tokyo.ac.jp (CMI)}
Kenta Hotokezaka$^{1}$
\\
$^{1}$Research Center for the Early Universe, Graduate School of Science, The University of Tokyo, Bunkyo, Tokyo 113-0033, Japan
}

\date{Accepted XXX. Received YYY; in original form ZZZ}

\pubyear{\the\year{}}

\begin{document}
\label{firstpage}
\pagerange{\pageref{firstpage}--\pageref{lastpage}}
\maketitle

\begin{abstract}
The spectrum of the first supernova light (i.e., the shock breakout and early cooling emission) is an important diagnostic for the state of the progenitor star just before explosion. We consider a streamlined model describing the emergent shock breakout spectrum, which enables a straightforward classification of the possible observed spectra during the early planar phase.  The overall spectral evolution is determined by a competition between three important time-scales: the diffusion time $t_{\rm{bo}}$ of the shell producing the breakout emission, the light-crossing time of this shell $t_{\rm{lc}}$, and the time $t_{\rm{eq}}$ at which the observer starts to see layers of the ejecta where the gas and radiation are in thermal equilibrium.  There are five allowed orderings of these time-scales, resulting in five possible scenarios with distinct spectral behaviours.  Within each scenario, the spectrum at a given time is one of five possible types, which are approximately described by broken power-laws; we provide the spectral and temporal indices of each power-law segment, and the time evolution of the break frequencies.  If high-cadence multi-wavelength observations can determine the relevant breakout scenario in future events, strong constraints can be placed on the physical conditions at the site of shock breakout.
\end{abstract}




\section{Introduction}
\label{sec:introduction}

Shock breakout emission is the first observable light from a supernova (SN), which is produced once the optical depth to the radiation-mediated SN shock becomes low enough that its radiation can leak out \citep{colgate,falk,klein}.  The initial burst of radiation is followed by a planar phase of evolution, lasting until the shock has doubled its radius since the breakout \citep[e.g.,][]{piro,ns,ns2,sapir1,katz2,fs}.  For a breakout radius of $R$ and a shock velocity of $v_{\rm{bo}}$, this planar phase lasts for a time of $t_{\rm{s}} \sim R/v_{\rm{bo}}$, with the value of $t_{\rm s}$ ranging from seconds for fast shocks in compact stars, to days for slow shocks in extended media.  Detecting the emission during this early phase is of considerable interest, as it potentially provides important information about the progenitor star and its immediate circumstellar environment.  It is therefore crucial to understand the multi-wavelength emission produced by shock breakout at these early times.

For slow shock velocities, the postshock gas and radiation are expected to be in thermal equilibrium, in which case the observed spectrum is well-described by a blackbody spectrum \citep[e.g.,][]{ns,fs,morag1,morag2,morag3}.  However, the assumption of equilibrium does not hold for sufficiently fast shocks with $v_{\rm{bo}} \ga 0.1\,c$.  Under standard assumptions, the emergent spectrum in the non-equilibrium case is generally a free-free spectrum, which is further altered by self-absorption and Comptonization effects \citep[e.g.,][]{weaver,katz,ns,sapir3,waxmankatz,fs}. Computing the spectrum in the non-equilibrium case is not straightforward, as the observed temperature depends on the complicated physics of thermalization and Comptonization in the breakout ejecta.  In a significant part of parameter space, the situation is further complicated by the non-negligible difference in arrival time of photons originating from different latitudes.  As a result, the literature covering the early spectrum tends to be technical, with individual works often restricted to particular cases of interest. Synthesizing existing work on the topic, and distilling it into a more accessible form, would therefore be of value to the broader transient community.  Our aim in this work is to take the first step towards this synthesis. 

In this letter, we present a simplified approach to the shock breakout problem, which absorbs much of the complex physics into readily observable quantities.  We then explore how our description leads to a convenient and intuitive way to classify the possible observed spectra.  We start with an overview of the spectral model in Section~\ref{sec:overview}, which introduces three important time-scales that influence the breakout spectrum.  Next, we consider how the ordering of these time-scales affects the overall spectral evolution in Section~\ref{sec:spectralregimes}.  The possible types of spectra are enumerated in Section~\ref{sec:spectraltypes}, where we also provide the temporal evolution of their critical break frequencies and spectral luminosities. Finally, we briefly examine what happens after the end of the planar phase in Section~\ref{sec:spherical}, before concluding in Section~\ref{sec:conclusions} with a discussion of observational prospects. This paper is part of a series of related papers; interested readers can find more details about the underlying spectral model in Irwin \& Hotokezaka 2024a (submitted, hereafter Paper I), and an application of the model to low-luminosity gamma-ray bursts in Irwin \& Hotokezaka 2024b (submitted).

\section{Model Overview}
\label{sec:overview}

We compute the spectrum using the model of Paper I, which describes the shock breakout in terms of the density $\rho_{\rm{bo}}$ and shock velocity $v_{\rm{bo}}$ at the breakout location, the radius $R$ at which the breakout occurs, and the power-law index $n$ describing the density profile at that location, i.e. ${\rho(r) \propto (R-r)^n}$ for $r\sim R$.\footnote{Alternatively, for an assumed density profile, the mass of the breakout medium, $M$, and the energy deposited in that medium, $E_0$, can be used instead of $\rho_{\rm{bo}}$ and $v_{\rm{bo}}$ (as discussed in Paper I).}   The model describes the spectral evolution during the planar phase, when $t \la t_{\rm{s}} = R/v_{\rm{bo}}$.  At times $t\ga t_{\rm{s}}$, we use the standard model of \citet{ns}. Our results are valid for spherical, non-relativistic shocks,\footnote{Non-spherical effects are discussed further in, e.g., \citet{matzner}, \citet{il}, \citet{linial}, and \citet{goldberg}, and relativistic effects in, e.g., \citet{katz}, \citet{ns2}, and \citet{fs2}.} with temperatures not exceeding $\sim 50\,$keV so that pair creation is negligible, and with little enough Comptonization that a prominent Wien peak is not expected (see Paper I for further discussion and justification).  Electron scattering is assumed to be the dominant opacity source. Additionally, the shell producing the breakout emission, which we refer to as the `breakout shell' or `breakout layer,' is assumed to be thin compared to $R$.  The model applies to breakout from the stellar surface, and also to breakout from an extended region (e.g., a circumstellar medium or Type IIb-like low-mass envelope), provided that the breakout occurs near the edge of that region (i.e., for the compact case of \citealt{ci11} or the edge-breakout case of \citealt{khatami}).  If there is sufficient material beyond the breakout location, a significantly different spectrum may be expected due to photoabsorption and Compton degradation \citep[e.g.,][]{ci12,margalit}, and narrow emission lines may also be present as observed in Type IIn, Ibn, and Icn SNe.

\begin{figure}
    \centering
    \includegraphics[width=\linewidth]{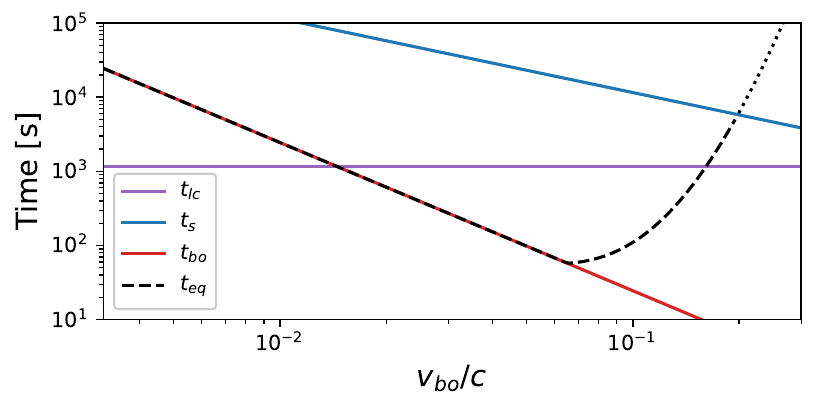}
    \caption{The time-scales $t_{\rm lc}$ (purple), $t_{\rm bo}$ (red), $t_{\rm eq}$ (black dashed), and $t_{\rm s}$ (blue) for $\rho_{\rm bo} = 10^{-9}\,\text{g}\,\text{cm}^{-3}$, $R = 500\,\mathrm{R}_\odot$, and $n=1.5$, assuming ${\kappa = 0.34\,\text{cm}^2\,\text{g}^{-1}}$.  For these parameters, we find that the breakout duration is dominated by the diffusion time for $v_{\rm bo} \la 0.015\,c$, and by the light-crossing time otherwise.  The breakout emission departs from thermal equilibrium above $v_{\rm bo}\ga 0.07\,c$.  In calculating $t_{\rm eq}$, we assumed $t_{\rm eq} < t_{\rm s}$; where this is not valid, the dashed line becomes dotted.} 
    \label{fig:tbo}
\end{figure}

In shock breakout, the gas and radiation in the breakout shell may or may not be in thermal equilibrium; this distinction strongly affects the observed temperature $T_{\rm{obs}}$.  In the equilibrium case, the spectrum is a blackbody spectrum with a slowly-evolving temperature $T_{\rm{BB}}$, and roughly ${T_{\rm{obs}} = T_{\rm{BB}} \propto t^{-1/3}}$.  In the non-equilibrium case, the spectrum is a Comptonized free-free spectrum with $T_{\rm{obs}} > T_{\rm{BB}}$, and $T_{\rm obs}$ evolves rapidly until deeper, thermalized layers are revealed \citep[e.g.,][]{fs}.  There are three relevant time-scales which shape the behaviour of the shock breakout spectrum:
\begin{itemize}
    \item The time-scale $t_{\rm{lc}} = R/c$, where $c$ is the speed of light, is the light-crossing time of the system.  It sets the maximum difference in light travel time for the radiation released during the breakout.
    \item The time-scale $t_{\rm{bo}}\approx c/\kappa\rho_{\rm bo} v_{\rm bo}^2$, where $\kappa$ is the electron-scattering opacity, is both the diffusion time of the breakout layer, and its dynamical time.  It determines the initial smearing of the signal by diffusion, and also sets the time-scale for  adiabatic cooling in the expanding flow. 
    \item The time-scale $t_{\rm{eq}}$ (see, e.g., Section 2 of Paper I for a derivation) is the time when ejecta in thermal equilibrium are revealed, which depends non-trivially on $\rho_{\rm bo}$, $v_{\rm bo}$, and $n$.  It satisfies $t_{\rm{eq}} = t_{\rm{bo}}$ if equilibrium holds in the breakout layer, and $t_{\rm{eq}} > t_{\rm{bo}}$ otherwise.
\end{itemize}
Note that our assumption of non-relativistic flow implies $v_{\rm bo} \ll c$ and $t_{\rm lc} \ll t_{\rm s}$, while our assumption that the breakout shell is thin implies $v_{\rm bo} t_{\rm bo} \ll R$ and $t_{\rm bo} \ll t_{\rm s}$.  In principal, $t_{\rm eq}$ can be larger or smaller than $t_{\rm s}$ (see, however, Section~\ref{sec:spherical}). To give an example of the general behaviour, the time-scales $t_{\rm lc}$, $t_{\rm bo}$, and $t_{\rm eq}$, as well as the time-scale $t_{\rm s}$ introduced in Section~\ref{sec:introduction}, are plotted as functions of $v_{\rm bo}$ in Fig.~\ref{fig:tbo}.

In addition, the spectrum is influenced by:
\begin{itemize}
    \item The bolometric luminosity produced in the breakout layer, $L_{\rm{bo}} \approx 4 \uppi R^2 \rho_{\rm bo} v_{\rm bo}^3$.
    \item The radiation temperature in the breakout layer, $T_{\rm{obs,bo}}$, whose value depends on whether equilibrium holds or not.  If $t_{\rm eq} = t_{\rm bo}$, it is given by $T_{\rm obs,bo}\approx (\rho_{\rm bo}v_{\rm bo}^2/a)^{1/4}$ where $a$ is the radiation constant, while if $t_{\rm eq} > t_{\rm bo}$, it typically depends only on $v_{\rm bo}$ and $n$, but in a complicated way (see, e.g., Section 4 of Paper I).
    \item The time-averaged power-law index,\footnote{In reality, the temperature evolution at early times is not a simple power-law; it also depends strongly on a logarithmic term \citep[see, e.g.,][and Paper I]{fs}.  For simplicity, we replace this behaviour by a power-law with the same average slope.  Our main conclusions are not affected by this adjustment.} $\alpha$, of the observed temperature $T_{\rm{obs}} \propto t^{-\alpha}$ between $t_{\rm{bo}}$ and $t_{\rm{eq}}$ (given in Section 2 of Paper I, and only relevant if $t_{\rm{eq}} > t_{\rm{bo}}$).  It is a function of $\rho_{\rm bo}$, $v_{\rm bo}$, and $n$.
\end{itemize}
The total energy radiated during the breakout pulse is ${E_{\rm bo} \approx L_{\rm bo} t_{\rm bo}}$, and this energy is spread over a duration of $t_{\rm{obs}} \approx \max(t_{\rm{bo}},t_{\rm{lc}})$, so the observed luminosity of the breakout is ${L_{\rm{obs}} \approx L_{\rm{bo}}(t_{\rm{bo}}/t_{\rm{obs}})}$.    After the initial pulse, the luminosity follows $L\propto t^{-4/3}$ \citep[e.g.,][]{ns,fs} for $t<t_{\rm s}$.

\begin{figure*}
    \centering
    \includegraphics[width=\hsize]{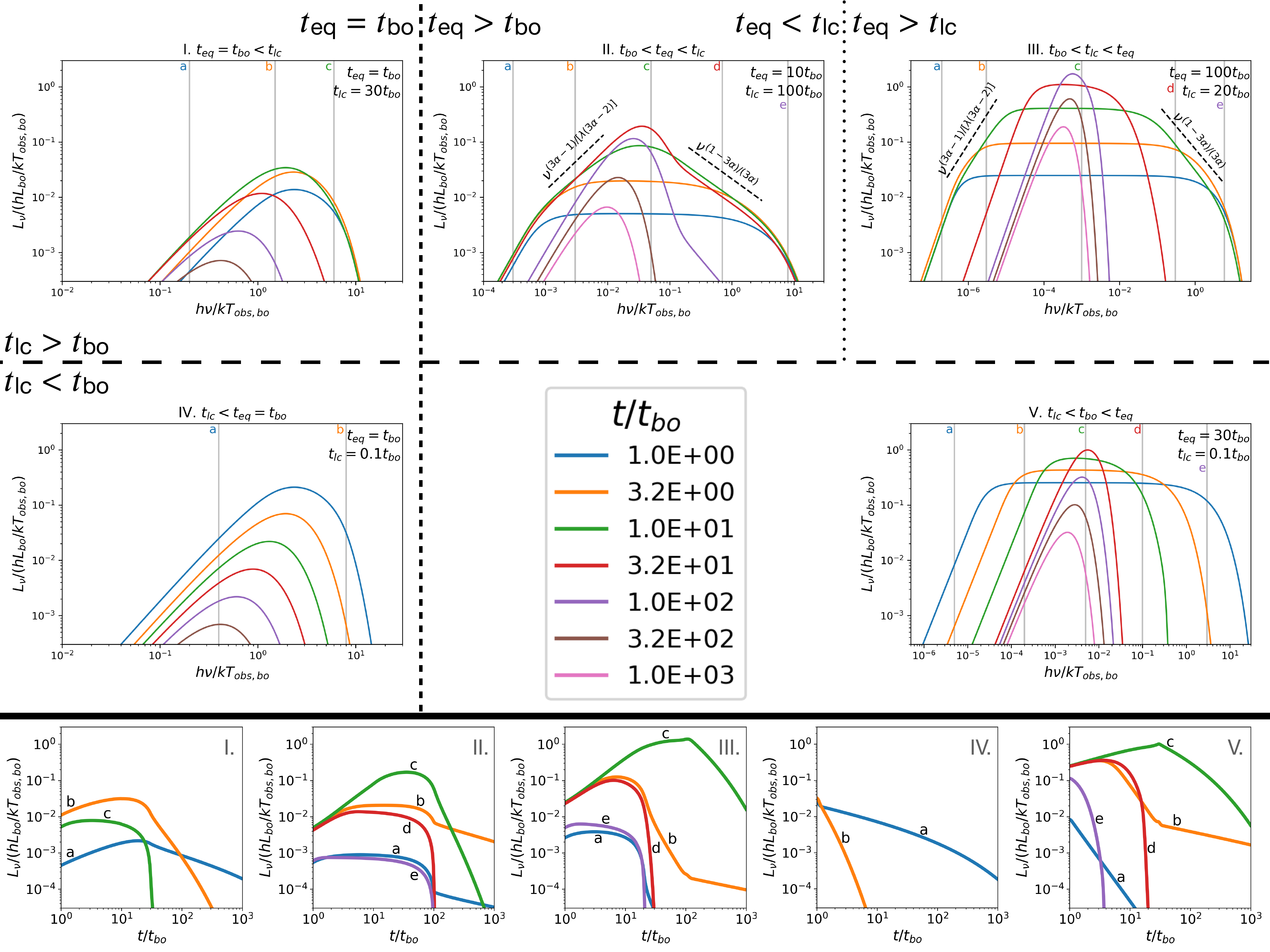}
    \caption{Example spectra (top) and light curves (bottom) for the five shock breakout scenarios discussed in Section~\ref{sec:spectralregimes}.   In the spectra, the line colour indicates time in units of $t_{\rm{bo}}$, as shown in the central legend.  To the right of the vertical dashed line, $t_{\rm{eq}} > t_{\rm{bo}}$ and the spectrum is a free-free spectrum at early times.  Above the long-dashed horizontal line, $t_{\rm{lc}} > t_{\rm{bo}}$ and the early-time spectrum is smeared out by the effects of light travel time; the black dashed bars in the upper middle and right panels indicate the approximate power-law indices resulting from this smearing.  The case where $\min(t_{\rm lc},t_{\rm eq}) > t_{\rm bo}$ is further subdivided by the dotted line, depending on whether or not $t_{\rm{eq}} < t_{\rm{lc}}$.  Select light curves (labelled with lower case letters) are shown for the frequencies indicated by the vertical gray lines (with matching coloured letters) in the spectral plots.}
    \label{fig:spectraevolution}
\end{figure*}

It is also useful to define three characteristic frequency scales: the breakout shell's initial peak frequency $\nu_{\rm{obs,bo}} \approx kT_{\rm{obs,bo}}/h$, its initial self-absorption frequency $\nu_{\rm{a,bo}}$, and the peak frequency at $t=t_{\rm{eq}}$, which is $\nu_{\rm{eq}} \approx kT_{\rm{eq}}/h$, where 
\begin{equation}
    T_{\rm{eq}} = T_{\rm{obs,bo}}(t_{\rm{eq}}/t_{\rm{bo}})^{-\alpha}
\end{equation} is the temperature of the first thermalized ejecta to be revealed.  In a non-equilibrium breakout, $\nu_{\rm{a,bo}} < \nu_{\rm{eq}} < \nu_{\rm{obs,bo}}$ initially. As time goes on, deeper ejecta layers with lower $T_{\rm{obs}}$ and higher $\nu_{\rm{a}}$ become visible, until eventually thermalized layers with $kT_{\rm{obs}} \approx h\nu_{\rm{a}} \approx kT_{\rm{eq}}$ are exposed at time $t_{\rm{eq}}$.

As alluded to in Section~\ref{sec:introduction}, our way of describing the problem glosses over the detailed physics of thermalization and Comptonization.  In other studies \citep[e.g.,][]{ns,fs}, thermal equilibrium is captured by a dimensionless parameter $\eta$ (where $\eta \le 1$ when thermal equilibrium holds, and $\eta > 1$ otherwise), while Comptonization is governed by a dimensionless parameter $\xi$ (where $\xi > 1$ when Comptonization is important and $\xi=1$ otherwise).  Our treatment is equivalent, but $\eta$ has effectively been absorbed into $t_{\rm{eq}}$, while $\xi$ has been absorbed into $T_{\rm{obs,bo}}$ and $\alpha$.

In summary, for times $t \la t_{\rm{s}}$, the observed bolometric luminosity of shock breakout obeys\footnote{Note that in the $t_{\rm lc} > t_{\rm bo}$ case, there is a sharp drop in luminosity when the last breakout emission arrives at $t \sim t_{\rm lc}$ \citep[e.g.,][]{il}.}
\begin{equation}
    \label{Lshort}
    L(t) \approx 
    \begin{cases}
        L_{\rm{bo}} \dfrac{t_{\rm{bo}}}{\max(t_{\rm{bo}},t_{\rm{lc}})}, & t<\max(t_{\rm{bo}},t_{\rm{lc}}) \\
        L_{\rm{bo}} \left(\dfrac{t}{t_{\rm{bo}}}\right)^{-4/3}, & t>\max(t_{\rm{bo}},t_{\rm{lc}})
    \end{cases},
\end{equation}
and the observed temperature follows
\begin{equation}
    \label{Tshort}
    T_{\rm{obs}}(t) \approx 
    \begin{cases}
    T_{\rm{eq}} \left(\dfrac{t}{t_{\rm{eq}}}\right)^{-\alpha}, & t<t_{\rm{eq}} \\
    T_{\rm{eq}} \left(\dfrac{t}{t_{\rm{eq}}}\right)^{-1/3}, & t>t_{\rm{eq}}
    \end{cases}.
\end{equation}
For $t<t_{\rm{eq}}$, the spectrum is a Comptonized free-free spectrum, self-absorbed below a frequency of
\begin{equation}
    \label{nuashort}
    \nu_{\rm{a}}(t) \approx 
    \dfrac{kT_{\rm{eq}}}{h}\left(\dfrac{t}{t_{\rm{eq}}}\right)^\beta,
\end{equation}
where to a reasonable approximation $\beta\approx \lambda(\alpha-2/3)$ with ${\lambda \approx 1.2\text{--}1.5}$ (Paper I). For $t>t_{\rm eq}$, the spectrum is a blackbody spectrum.

As a final note, if light travel time is important ($t_{\rm{lc}} > t_{\rm{bo}}$), then due to the spread in arrival times across the breakout surface, emission with a range of temperatures and self-absorption frequencies is simultaneously observed for $t<t_{\rm{lc}}$.  The observed temperatures lie in the range $T_{\rm{obs}}(t)$ to $T_{\rm{obs,bo}}$, and the self-absorption frequencies in the range $\nu_{\rm{a,bo}}$ to $\nu_{\rm{a}}(t)$.  Within these frequency ranges a power-law behaviour is expected \citep[Paper I; see also][]{ns,fs}.

With good enough wavelength coverage at early enough times, all of the quantities introduced above can in principle be probed by observations.  The temperature, self-absorption frequency, and luminosity could be inferred directly from the SED. The rise time of the bolometric luminosity gives $t_{\rm bo}$, and the duration of the breakout signal gives the longer of $t_{\rm{bo}}$ and $t_{\rm{lc}}$.  The $t_{\rm lc} < t_{\rm bo}$ and $t_{\rm bo} < t_{\rm lc}$ cases could therefore be distinguished by whether the rise time is comparable to, or shorter than, the duration. The value of $t_{\rm{eq}}$ could be measured by determining the time when a blackbody emission component first appears in the spectrum, with the blackbody temperature at that time corresponding to $T_{\rm{eq}}$.

\section{The five possible breakout scenarios}
\label{sec:spectralregimes}

There are six possible permutations of the time-scales $t_{\rm{bo}}$, $t_{\rm{eq}}$, and $t_{\rm{lc}}$.  However, the condition $t_{\rm{eq}} \ge t_{\rm{bo}}$ rules out the case where ${t_{\rm{eq}} < t_{\rm{lc}} < t_{\rm{bo}}}$.  Each of the five remaining orderings results in a different breakout scenario with a distinct spectral behaviour.  We label these scenarios with capital Roman numerals I--V. 

\begin{figure*}
    \centering
    \includegraphics[width=\textwidth]{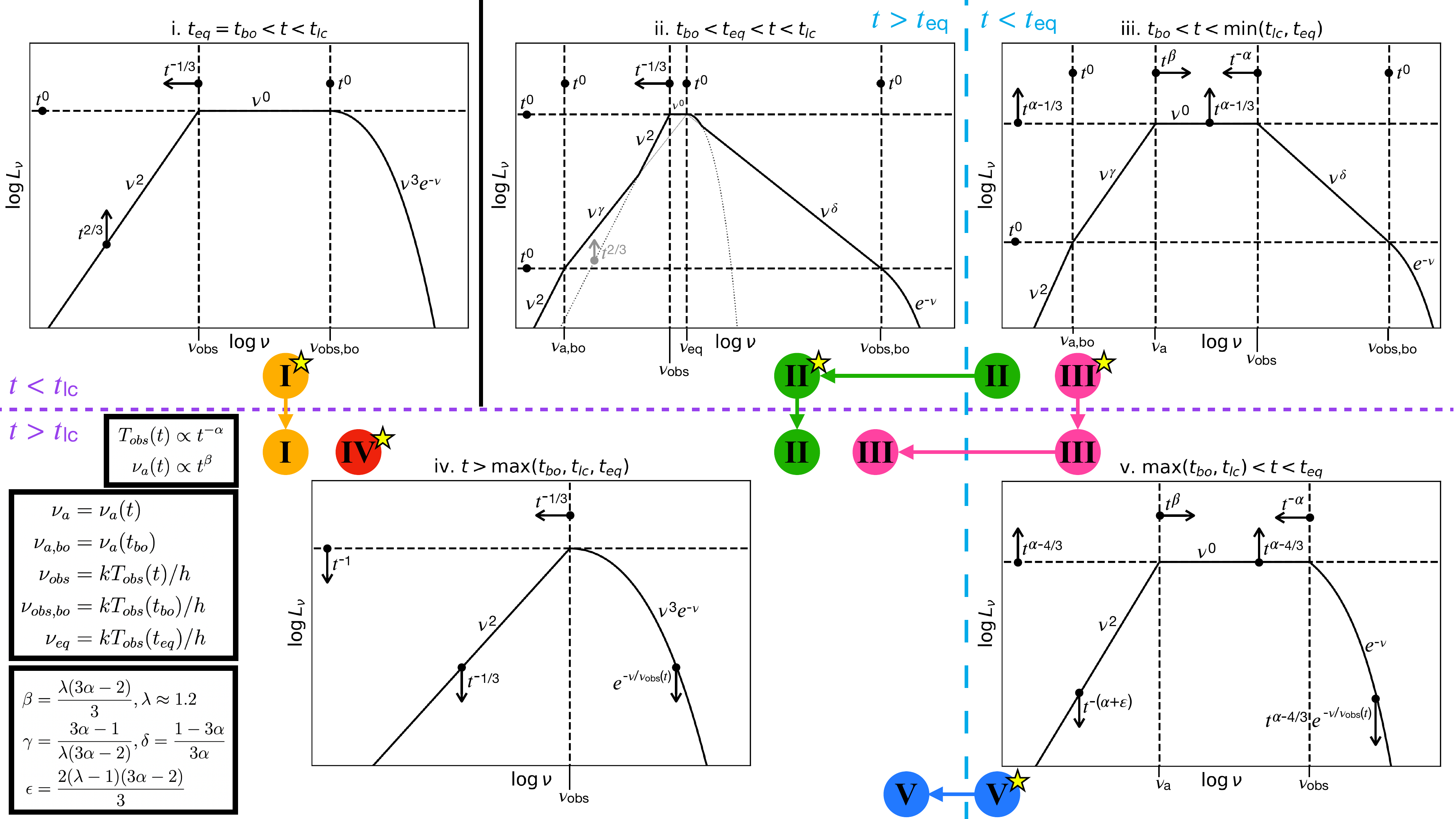}
    \caption{The five possible types of spectra discussed in Section~\ref{sec:spectraltypes}. In the top middle panel, the separate contributions from free-free (solid gray) and blackbody (dotted gray) emission are also shown.  The long-dashed light blue line separates the spectra with significant blackbody emission ($t> t_{\rm{eq}}$) from those without it ($t<t_{\rm{eq}}$). Likewise the short-dashed purple line separates spectra which are smeared due to a significant spread in light arrival time ($t <t_{\rm{lc}}$) from those which are not ($t > t_{\rm{lc}})$.  In each case, the relevant break frequencies $\nu_{\rm{obs,bo}}$, $\nu_{\rm{obs}}$, $\nu_{\rm{eq}}$, $\nu_{\rm{a}}$, and $\nu_{\rm{a,bo}}$ are marked with dashed vertical lines, and the corresponding values of $L_\nu$ with dashed horizontal lines, with each line labelled according to its time dependence.  For each segment of the spectrum, the spectral index is also shown, along with an arrow indicating the time evolution (if non-constant). The quantities $\beta$, $\gamma$, $\delta$, and $\epsilon$ are functions of $\alpha$ as given in the lower-left box.   The coloured tokens and arrows show how the spectral type evolves over time in each of the scenarios discussed in Section~\ref{sec:spectralregimes}. A yellow star next to a token denotes the expected spectral type at peak bolometric light.}
    \label{fig:spectratypes}
\end{figure*}

The evolution of the spectrum in each scenario, as calculated using the model of Paper I, is shown in Fig.~\ref{fig:spectraevolution}. Briefly, the behaviour in each case is as follows:
\begin{itemize}
\item Scenario I ($t_{\rm{eq}} = t_{\rm{bo}} < t_{\rm{lc}}$):  The spectrum is a blackbody spectrum, slightly broadened due to the non-negligible light-crossing time.  The spectral luminosity $L_\nu$ remains constant until $t\approx t_{\rm{lc}}$, then starts declining.
\item Scenario II ($t_{\rm{bo}} < t_{\rm{eq}} < t_{\rm{lc}}$): The spectrum is initially a self-absorbed free-free spectrum.  For $\alpha > 4/3$, $L/T_{\rm{obs}}$ is initially increasing, so the peak $L_\nu$ rises.  As time goes on, the spread in light arrival time smears the spectrum into a broken power-law.  A blackbody component develops once thermalized ejecta are exposed at $t=t_{\rm{eq}}$.  At this time the breakout is still ongoing and free-free emission is also present.  After the breakout flash ends at $t=t_{\rm{lc}}$, the free-free emission fades and the spectrum becomes a blackbody spectrum.
\item Scenario III ($t_{\rm{bo}} < t_{\rm{lc}} < t_{\rm{eq}}$): The early evolution is the same as in scenario II, but when the breakout ends at $t=t_{\rm{lc}}$, no thermalized ejecta have become visible yet.  Since light travel time is no longer important once $t>t_{\rm{lc}}$, an ordinary (i.e., not smeared) free-free spectrum is observed for $t_{\rm lc} < t < t_{\rm{eq}}$, and after that the spectrum becomes a fading blackbody spectrum.
\item Scenario IV ($t_{\rm{lc}}$ < $t_{\rm{eq}} = t_{\rm{bo}}$): The spectrum is a cooling and fading blackbody spectrum throughout the entire evolution. 
\item Scenario V ($t_{\rm{lc}} < t_{\rm{bo}} < t_{\rm{eq}}$): The spectrum is initially an ordinary free-free spectrum, unaffected by light travel time.  As with Scenarios II and III, the peak $L_\nu$ rises at early times, albeit more slowly.  Once thermal ejecta are revealed at $t = t_{\rm{eq}}$, the spectrum becomes a blackbody spectrum and it begins to fade away.
\end{itemize}  
The relatively complicated physics in scenario II are discussed extensively in Paper I, and applied to low-luminosity GRBs in a separate paper (Irwin \& Hotokezaka 2024b).

Which of the above scenarios is relevant depends on the initial conditions (i.e., on $\rho_{\rm{bo}}$, $v_{\rm{bo}}$, $R$, and $n$).  For example, in Fig.~\ref{fig:tbo}, we see that as $v_{\rm bo}$ is increased, the appropriate scenario changes from IV to I, then to II, and finally to III.  If a lower density or radius were chosen, so that the $t_{\rm eq}$ and $t_{\rm lc}$ lines did not intersect, the progression would instead be from IV to V to III with increasing $v_{\rm bo}$ (see also, e.g., Section 3 of Paper I). Therefore, if early multi-wavelength observations can distinguish between these scenarios, meaningful constraints can be placed on the properties of the progenitor and the explosion. Often, a determination of $t_{\rm lc}$, $t_{\rm bo}$, and the total radiated energy $E_{\rm bo}$ is sufficient to estimate the radius, density, and velocity of the breakout shell.  In scenarios I--III, where $t_{\rm lc} > t_{\rm bo}$, we have $R \approx c t_{\rm lc}$, $\rho_{\rm bo} \approx 16\uppi^2 c^7t_{\rm lc}^4/\kappa^3 t_{\rm bo} E_{\rm bo}^2$, and $v_{\rm bo} \approx \kappa E_{\rm bo}/4 \uppi c^3 t_{\rm lc}^2$.  In scenarios IV and V, however, $t_{\rm lc}$ cannot be determined, so only the degenerate quantities $\rho_{\rm bo} v_{\rm bo}^2 \approx c/\kappa t_{\rm bo}$ and $R^2 v_{\rm bo} \approx \kappa E_{\rm bo} /4 \uppi c$ are constrained.  In scenario V the degeneracy can be broken by using the observed temperature $T_{\rm obs,bo}$ to get $v_{\rm bo}$ independently (see, e.g., Section 4 of Paper I), but in scenario IV the degeneracy is inherent since in that case $T_{\rm obs,bo}$ and $t_{\rm bo}$ both depend only on the quantity $\rho_{\rm bo}v_{\rm bo}^2.$  The determination of $n$ is not straightforward and is beyond the scope of this paper.

\section{The five fundamental types of breakout spectra}
\label{sec:spectraltypes}

In any of the breakout scenarios I-V discussed in Section~\ref{sec:spectralregimes}, the spectrum at a given time is  one of five fundamental types.  We label each of these types of spectra with lower-case Roman numerals i--v.  The relevant type depends on how the time $t$ compares to the characteristic time-scales $t_{\rm{bo}}$, $t_{\rm{lc}}$, and $t_{\rm{eq}}$.  The five possibilities are:
\begin{itemize}
\item Type i ($t_{\rm{eq}} = t_{\rm{bo}} < t < t_{\rm{lc}}$): A blackbody spectrum, smeared out by the effects of light travel time. 
\item Type ii ($t_{\rm{bo}} < t_{\rm{eq}} < t < t_{\rm{lc}}$): A superposition of smeared-out free-free and blackbody components.
\item Type iii ($t_{\rm{bo}} < t < \min(t_{\rm{lc}},t_{\rm{eq}})$): A self-absorbed free-free spectrum, smeared out by the effects of light travel time.
\item Type iv ($t > \max(t_{\rm{bo}},t_{\rm{lc}},t_{\rm{eq}})$): A blackbody spectrum.
\item Type v ($\max(t_{\rm{bo}},t_{\rm{lc}}) < t < t_{\rm{eq}}$): A self-absorbed free-free spectrum.
\end{itemize}
Near peak light, i.e., when $t$ is an appreciable fraction of ${t_{\rm{obs}} \approx \max(t_{\rm{lc}},t_{\rm{bo}})}$, the spectrum in each scenario is of the type with the same Roman numeral (i.e., in scenario I, the spectrum at peak light is of type i, and so on).

To further clarify the situation and our nomenclature, the relevant \textit{scenario} depends on the \textit{initial conditions} of the breakout; this sets the ordering of the three characteristic time-scales and determines which of the spectral types are permissible.  Then, within a given scenario, as $t$ increases from an initial value of $t_{\rm{bo}}$, the spectrum evolves through a prescribed sequence of the allowed spectral \textit{types}, with the observed spectrum shifting to a new type each time $t$ becomes equal to one of the characteristic time-scales.  The five fundamental spectral types can be thought of as `building blocks,' with the five breakout scenarios being constructed by stringing together these building blocks in different orders.

The main result of this study is presented in Fig.~\ref{fig:spectratypes}, where we show each of the five spectral types, how the characteristic frequencies of each type vary with time, and which of the types are expected in each of the five scenarios discussed above.  To the extent that our assumptions hold,\footnote{As discussed above, asphericity, relativistic effects, dense material outside the breakout shell, and significant Comptonization can all markedly alter the observed spectrum.} a shock breakout spectrum obtained at any time $t\la t_{\rm{s}}$ should be expected to roughly follow one of these five behaviours.  The evolution of the characteristic frequencies follows from equations~\ref{Tshort} and~\ref{nuashort}, and the bolometric luminosity evolves according to equation~\ref{Lshort}.  The value of $L_\nu$ always peaks at $\nu_{\rm{obs}} \approx kT_{\rm{obs}}(t)/h$, while the frequency which maximizes $\nu L_\nu$ (i.e., the peak energy) is either $\nu_{\rm{obs,bo}}$ (if $t<t_{\rm{lc}})$, or $\nu_{\rm{obs}}$ (if $t>t_{\rm{lc}}$).  A derivation of the spectral indices for the case of a smeared free-free spectrum is given in Paper I \citep[see also, e.g.,][]{ns,fs}.  The temporal evolution of $L_\nu$ follows from the known spectral indices and break frequencies, along with fact that the maximum value of $\nu L_\nu$ must track the bolometric luminosity.

\section{The spherical phase}
\label{sec:spherical}

So far, we have only considered the evolution in the planar phase, when $t<t_{\rm{s}}$.  Once $t > t_{\rm{s}}$, the optical depth of the ejecta starts to drop appreciably and deeper layers of the ejecta are more readily revealed.  Within a few $t_{\rm{s}}$, thermalized ejecta will be exposed, if they were not already \citep[e.g.,][]{ns,fs}.  Therefore, regardless of the conditions at early times, a blackbody spectrum is expected throughout most of the spherical phase.  In other words, $t_{\rm eq}$ is at most a few $t_{\rm s}$. As shown by \citet{ns}, the temperature for $t>t_{\rm{s}}$ evolves as $T_{\rm{obs}} \propto t^{-0.6}$, with a typically weak dependence on $n$, while the luminosity evolution is somewhat more sensitive to $n$: ${L \propto t^\omega}$ with ${\omega = -(2.28n-2)/[3(1.19n+1)]}$.  For $n=1.5$--3, ${L\propto t^{-0.2}\text{--}t^{-0.4}}$, the peak value of $L_\nu$ follows ${L_\nu \propto L/T_{\rm{obs}} \propto t^{0.2}\text{--}t^{0.4}}$, and the light curve at frequencies $h\nu < kT_{\rm{obs}}$ obeys $L_\nu \propto L/T_{\rm{obs}}^3 \propto t^{1.4}\text{--}t^{1.6}$.

\section{Conclusions}
\label{sec:conclusions}

We have considered a very general model for the spectra produced by planar shock breakout, and have introduced a conceptually simple, observationally motivated parametrization of the problem which enables a straightforward classification of the possible spectral behaviours.  We found that there are five possible types of spectra which can be exhibited by planar shock breakout (as shown in Fig.~\ref{fig:spectratypes}), and gave the approximate scalings of the critical frequencies and spectral luminosities with time in each case.  In addition, we showed that the overall spectral evolution follows one of five possible scenarios (as shown in Fig.~\ref{fig:spectraevolution}), each of which evolve through a different sequence of spectral types over time.  Each scenario corresponds to one of the five allowed orderings of three characteristic time-scales: the initial diffusion time through the breakout layer, $t_{\rm{bo}}$; the light-crossing time, $t_{\rm{lc}}$; and the time $t_{\rm{eq}}$ at which thermalized ejecta are first revealed.

The spectral model presented here can also be used to generate multi-band light curves, as demonstrated in Fig.~\ref{fig:spectraevolution}.  A wide variety of diverse light curves are possible, depending on the relevant scenario and on how the observed frequency compares with the characteristic frequencies in Fig.~\ref{fig:spectratypes}.  For brevity, we do not consider this menagerie of light curve behaviours here, although we can identify some basic trends, like the achromatic sharp drop at $t \approx t_{\rm lc}$ exhibited when the breakout time-scale is dominated by the light-crossing time.  Classification of the multi-band light curves and photometric predictions for various progenitor systems will be pursued in subsequent work.

Although there has not yet been an event with sufficiently early multi-wavelength observations to put the planar shock breakout model to the test in the non-equilibrium case, we expect this situation to change in the near future, as current and future high-cadence facilities continue to push our observational capabilities to ever-shorter time-scales.  Our understanding is currently limited by a lack of coverage in the UV, as emission at UV wavelengths is a critical probe of shock breakout physics.  Thankfully, upcoming missions such as \textit{ULTRASAT} and \textit{UVEX} will alleviate this issue, and \textit{Einstein Probe} will also be invaluable due to its sensitivity and wide field of view in soft X-rays.  While these satellites will be excellent for discovering shock breakout candidates, prompt ground-based follow-up in the optical and IR is also vital to fully characterize the spectrum.  We eagerly anticipate the new insights on shock breakout that upcoming facilities will undoubtedly provide.

\section*{Acknowledgements}

This work is supported by the JST FOREST Program (JPMJFR2136) and the JSPS Grant-in-Aid for Scientific Research (20H05639, 20H00158, 23H01169, 23H04900). 

\section*{Data Availability}


The data underlying this article will be shared on reasonable request
to the corresponding authors.



\bibliographystyle{mnras}








\bsp	
\label{lastpage}
\end{document}